\documentclass[preprint,12pt]{elsarticle}

\usepackage{graphicx}
\usepackage{epsfig}

\usepackage{amssymb}

\journal{Physics Letters B}

\begin{document}

\begin{frontmatter}



\title{\vskip -2.5cm \hfill BI-TP 2010/03, TIFR/TH/10-02 \vskip 2.0 cm
Renormalized Polyakov loop in the Fixed Scale Approach}

\author[label1,label2]{Rajiv V. Gavai\corref{cor1}}
\ead{gavai@tifr.res.in}
\address[label1]{Fakult\"at f\"ur Physik, Universit\"at Bielefeld,
D-33615 Bielefeld, Germany}
\address[label2]{Department of Theoretical Physics, Tata Institute of
Fundamental Research, \\ Homi Bhabha Road, Mumbai 400005, India}
\cortext[cor1]{On sabbatical leave from
Tata Institute of Fundamental Research, Mumbai, India.}

\begin{abstract}

I compute the deconfinement  order parameter for the $SU(2)$ lattice gauge
theory, the Polyakov loop, using the fixed scale approach for two different
scales and show how one can obtain a physical, renormalized, order parameter.
The generalization to other gauge theories, including quenched or full
QCD, is straightforward. 

\end{abstract}


\end{frontmatter}



Since the pioneering work of Hagedorn \cite{hag} on the limiting
temperature for hadrons, it has been widely expected that the strongly
interacting matter will show unusual features at high enough
temperatures.   With the advent of quantum chromodyanmics (QCD) as the theory
of strong interaction with quarks and gluons as its basic constituents,
quark-gluon plasma was identified as this new phase with a possible phase
transition in between.  Using the lattice formulation of QCD, such a transition
was shown to be akin to spin models.  The Polyakov loop, $L$, defined as the
product of the timelike gauge links at a given site, is the order parameter for
this deconfinement transition \cite{McL}.  
On an Euclidean $N_\sigma^3 \times N_\tau$ lattice $L(\vec x)$ is defined 
at a site $\vec x$ as
\begin{equation}
L(\vec x) = \frac {1}{N_c}~{\rm Tr}~\Pi^{N_\tau}_{x_0=1}~U^4(\vec x, x_0)~,~
\end{equation}
where $N_c$ is the number of colours, being three for QCD, and
$U^\mu(x)$ are the gauge variables associated with the directed links in
the $\mu$th direction, $\mu=1$,4.  It is convenient to define its average over 
the spatial volume, $\bar L = \sum_{\vec x} L(\vec x) / N^3_\sigma$. 
$\langle |\bar L| \rangle$ was used to establish a second 
order deconfinement transition in numerical simulations of the $SU(2)$ pure 
gauge theory.  Since then it has been used for similar studies of the 
deconfinement phase transitions for a variety of $N_c$ \cite{Cel}, for 
establishing the universality \cite{Gav} of the continuum limit, as well as 
for theories with dynamical quarks \cite{DeT}.   Indeed, one hopes to be 
able to construct effective actions \cite{Pisa} for it in a Wilsonian RG 
approach, which will be similar to the spin models in the same universality 
class but with possibly additional interaction terms.  A large number of 
models of quark-hadron transitions use the Polyakov loop as the order 
parameter for the deconfinement transition as well.

An order parameter should be physical, i.e., independent of the 
lattice spacing $a$ in the continuum limit.  Furthermore, it must be so
in {\em both} the phases it seeks to distinguish.  As is the case for any
bare Wilson loop, the Polyakov loop, needs to be renormalized for this to
be true.  Since the bare Polyakov loop is further known to decrease 
progressively with $N_\tau$, suggesting it to be zero in the
continuum limit in the high temperature phase, renormalized $L$ is
even more desirable to have.

\begin{figure}
\begin{center}
   \scalebox{0.8}{\includegraphics{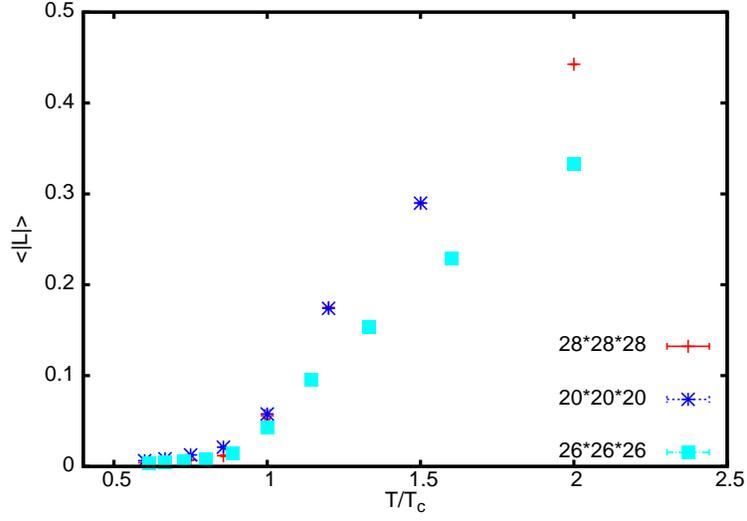}}
\end{center}
\caption{The average Polyakov loop as a function of $T/T_c$ for two different 
scales.   The lattice sizes are as indicated in the key.  }
\label{fg.l68}\end{figure}

\begin{figure}
\begin{center}
   \scalebox{0.8}{\includegraphics{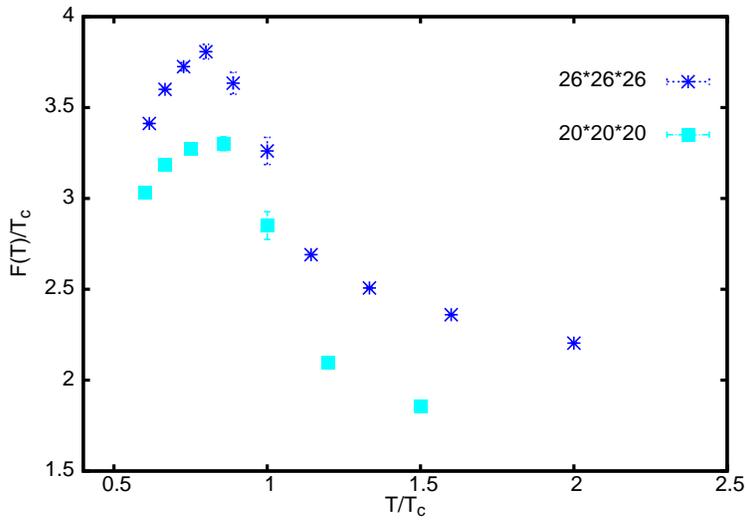}}
\end{center}
\caption{The heavy quark free energy $F$ a function of $T/T_c$ for two 
different scales.   The lattice sizes are as indicated in the key.  }
\label{fg.bf68}\end{figure}

The physical interpretation of the order parameter as a measure of the free
energy of a single quark, $\langle \bar L (T) \rangle = \exp ( -F_Q(T)/T) $
provides a  straightforward clue for renormalization.  Since many years it
is known that the single quark free energy contains divergent contribution
in the continuum limit, as shown by the computations employing  lattice
perturbation theory \cite{Hel}.  Subtracting these off was the first step
\cite{Hel} towards the renormalized Polyakov loop.
Kaczmarek et al. \cite{olaf1} proposed to use the heavy quark-antiquark
free energy, as determined from the Polyakov loop correlations at short
distances to  define the renormalized the Polyakov loop and showed it to
then become $N_\tau$-independent. Subsequently, fits to $\langle \bar L
\rangle$ on $N_\tau$-grids \cite{Dumi} and an iterative direct
renormalization procedure \cite{olaf2} for $\langle \bar L \rangle$
were used to extract the renormalize Polyakov loop, and shown to yield
similar results. 

While these are clearly nice results, it would be more satisfying to 
have a better definition of the renormalized order parameter for the following
reasons.  The definition of Ref. \cite{olaf1} needs heavy quark potential
at short distances.  The lattice artifacts are at their worse when one is at the
shortest distance, with maximal violation of the rotational invariance.
Finite volume of the lattice also enters in defining the maximum
distance between the heavy quarks, or Polyakov loops.  Similarly the
iterative procedure used in Ref. \cite{olaf2} to obtain the
renormalization constants needs large lattices in both spatial and
temporal directions, and should ideally be further tested on lattice of another
temporal size to check whether the same renormalization constants
apply.  Physically perhaps an undesirable aspect of the definition of 
Ref. \cite{olaf2} is that it works only on the plasma side, i.e., for 
$T \ge T_c$, where $T_c$ is the position of the peak in the Polyakov loop 
susceptibility.  The other definition \cite{olaf1} has so far been employed 
only in the $T \ge T_c$ for pure gauge theories for which $L$ is an
order parameter.  Furthermore, it would be nice if the renormalization
procedure is applicable to the usually employed $\langle |\bar L|
\rangle$, which is used as an order parameter  on finite volumes.

In this letter I show that a renormalized Polyakov loop which is valid for 
both the phases below and above $T_c$ can be defined, and it becomes 
the true order parameter in the infinite spatial volume limit.  Of
course, it is also physical, i.e., $N_\tau$-independent on finite
volumes as well.  I use the fixed scale approach
\cite{whot} to do so. It was introduced to minimize the computational
costs for the zero temperature simulations needed to subtract the vacuum
contribution in thermodynamic quantities such as the pressure and to
isolate pure thermal effects in computation of $T_c$ \cite{whotlat08}.
Furthermore, its advantage is that all the simulations stay on the line of 
constant physics in a straightforward way.  What I argue is that it 
is indeed this advantage which also permits an easy renormalization of the
Polyakov loop.  Although these considerations are general, and apply to any 
$SU(N)$ gauge theory, I shall consider below the simplest case of the
$SU(2)$ lattice gauge theory to illustrate how and why it works.

Recall that the temperature $T$ is varied in this approach by varying
$N_\tau$, holding the lattice spacing $a$, or equivalently the gauge coupling
$\beta = 2N_c/g^2$ fixed.  The single quark free energy $F_b(N_\tau, a)$ is 
then obtained from the $\bar L$ by the canonical relation, 
\begin{equation}
\ln \langle | \bar L | \rangle = -  a N_\tau F_b(N_\tau,a)~.
\label{eq.fe}
\end{equation}
The subscript $b$ reminds us that one obtains the bare free energy this
way.   If the chosen coupling is $\beta_c$, corresponding to the position
of the peak of the $|L|$-susceptibility in the usual fixed $N_\tau$
approach, and it lies in the scaling region, then the physical
deconfinement temperature $T_c = 1/N_{\tau,c} a_c$, and in the fixed scale
approach then $T/T_c$ = $N_{\tau,c}/N_\tau$, with the free energy given by
$F_b(T/T_c, a_c)$.  Writing it as a sum of a divergent and a regular
contribution, one has $a_c F_b(T/T_c, a_c) = a_c F(T/T_c,a_c) - a_c
A(a_c)$, where $A$ is the divergent free energy in physical units.
Clearly, the  divergent contribution will be {\em same} at all temperatures
in the fixed scale approach since it depends only on $a_c$.   

Since $\beta_c$, or $a_c$, is known precisely for the Wilson action
of the $SU(2)$ theory for many different $N_\tau$, I chose $\beta_{c1} 
= 2.4265$, and $\beta_{c2} = 2.5104$ corresponding to the known transition 
temperatures on $N_\tau = 6$ \cite{Eng} and 8 \cite{Vel} respectively.  
Note that $T/T_c$ is given 
simply by $6/N_\tau$ and $8/N_\tau$ respectively.  Employing then $N_\tau = 3$
to 12, I varied the temperature in the range 2 $\ge T/T_c \ge 0.6$.
I used two different spatial lattice sizes for the smaller $\beta$,
$20^3$ and $28^3$, while a $26^3$ lattice was used for the larger one.
Note that fixed $a_c$ means that the spatial volume was constant in physical
units in each case, being 37.03 and 101.63 in the units of $T_c^{-3}$
for $\beta_{c1}$ and 34.44 for $\beta_{c2}$.  Note that in contrast to
these simulations,  the spatial volume varies with $T$ in the usual fixed 
$N_\tau$ approach.
 
Figure \ref{fg.l68} shows the results for the thermal expectation value of
$\bar L $ as a function of the temperature in the units of $T_c$. In most
cases, I used both a random and an ordered start.  The errors are corrected
for autocorrelations.  The agreement in the data for the two starts suggest
the statistics of 200K iterations to be sufficient.   As expected, the two
different scales lead to two different curves for the order parameter.
Figure \ref{fg.bf68} displays the behaviour of the corresponding bare free
energy, obtained by using the eq.(\ref{eq.fe}). It reinforces the
expectation of the effect of the divergent free energy, since it increases
with the decrease in the lattice cut-off $a_c$.

\begin{figure}
\begin{center}
   \scalebox{0.8}{\includegraphics{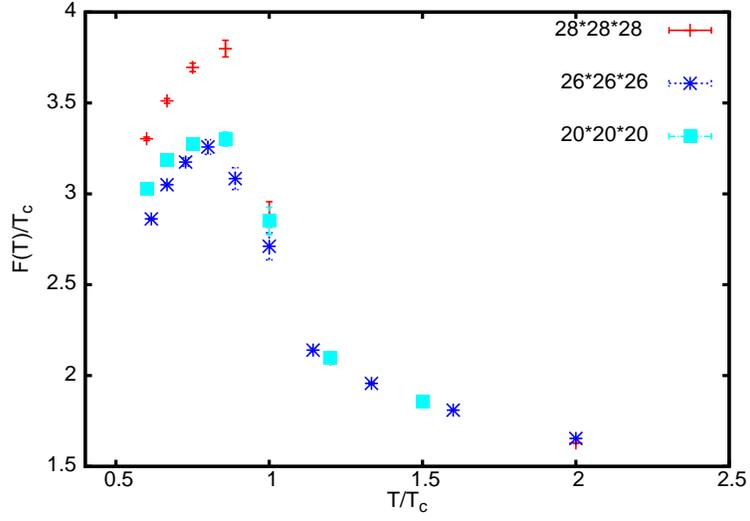}}
\end{center}
\caption{The same as Figure \ref{fg.bf68} but with a constant shift, as
explained in the text. }
\label{fg.f68}\end{figure}

\begin{figure}
\begin{center}
   \scalebox{0.8}{\includegraphics{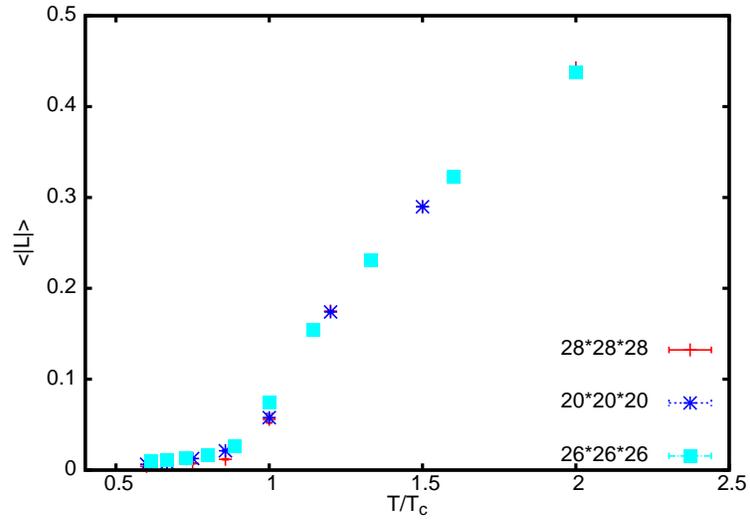}}
\end{center}
\caption{Same as Figure \ref{fg.l68} but with the shifted free energy
for the upper curve, as explained in the text.}
\label{fg.renl}\end{figure}

The two different scales, $a_{c1}$ and $a_{c2}$ have their respective
divergent contributions, $a_{c1} A(a_{c1})$ and $a_{c2} A(a_{c2})$.
Multiplying eq.(\ref{eq.fe}) by $N_j$, for $j=1$
and 2 corresponding to the critical $N_\tau$ for the scale choices
above, i.e, 6 and 8, one obtains
\begin{equation}
\frac{T}{T_c}  \ln \langle | \bar L | \rangle = -  \frac
{F_b(T/T_c, a_{cj})}{T_c}~,~
\label{eq.fe1}
\end{equation}
where $F_b(T/T_c, a_{cj})/T_c = F(T/T_c,a_{cj})/T_c -  A(a_{cj})/T_c$
Thus the free energies at the same temperatures but two different scales
are related by a mere constant, $[A(a_{c1})-A(a_{c2})]/T_c$.
Figure \ref{fg.f68} shows the same results as Figure
\ref{fg.bf68} but with a constant shift of -0.55 in the free energy for
the higher $\beta$.   A universal curve for the free energy seems to
result as a result for a wide range of $T/T_c$, covering both the
phases.  The finite free energy in the confined phase should not
surprise us.  In the
infinite volume limit, the free energy should increase to infinity in
the confined phase whereas it should essentially remain constant in the
deconfined phase. Such an expectation is indeed borne out by my results
for larger spatial volume, $28^3$ and $\beta_{c1}$, shown
in the same Figure \ref{fg.f68}.  One clearly sees an substantial increase 
in the free energy in the confined phase due to the 2.74 fold increase in 
spatial volume without affecting the high temperature phase in any significant 
way.  The points at the lowest $T/T_c$ seem to suggest a drop in the free 
energy in all cases, which I believe is a finite volume effect.  This is
also evident from the results for these points for the $28^3$ lattice 
in the same figure.  

Finally, it should now be clear how one can obtain a universal curve for 
the order parameter from the universal free energy curve.  
The $\langle | \bar L | \rangle $ obtained at the two 
different scales  will lie on a universal curve by simply multiplying 
the results for the scale corresponding to $\beta_{c2}$ by the
factor $\exp(N_\tau[A(a_{c1})/T_c -A(a_{c2})/T_c])$.  This is exhibited
in Figure \ref{fg.renl}.  This can be continued to as many scales as one 
wishes, and the same universal order parameter should result in {\em both} 
below and above $T_c$.  Furthermore, each new scale introduces only one
unknown constant which can be fixed by free energy difference at any $T
 > T_c$.  The entire low and high temperature region of the order
parameter is then uniquely fixed, and has to be universal.
As in case of any renormalized quantity, it
depends on the scale chosen to define the scheme.  Here it is the
inclusion of a constant free energy $A(a_{c})/T_c$ in the free energy 
for the chosen scale $a_c$ which defines the choice.  The details of the
shape of the physical order parameter are therefore scale-dependent in
the plasma phase but it is universal none the less once a choice is
made.  Moreover, any further change of scale leads to a computable
change in the shape.

In conclusion, I showed that the fixed scale approach leads to a natural 
definition of a physical, $N_\tau$-independent, order parameter which is 
defined in both the confined and the deconfined phases. The definition
itself does not depend on any lattice artifacts or the lattice size in the 
deconfined phase.  Moreover, it displays the expected behaviour in the
confined phase as the physical volume is increased, suggesting that the so 
determined physical free energy of a single quark in the confined phase, 
$F$, goes to infinity in the infinite volume limit.  It is
straightforward to generalize this idea to $SU(N_c)$ gauge theories and
QCD as well as to sources in higher representations.

This work was done during a visit to the University of Bielefeld and was
supported by the Deutsche Forschungsgemainschaft under the grant
GRK 881.  It is a great pleasure to acknowledge the kind hospitality 
of the Fakult\"at f\"ur Physik there and, in particular, that of Frithjof 
Karsch during the visit.




\bibliographystyle{elsarticle-num}
\bibliography{<your-bib-database>}

\begin{thebibliography}{00}

\bibitem{hag}
R.~Hagedorn,
  Nuovo Cim.\ Suppl.\  {\bf 3} (1965) 147.

\bibitem{McL}
  L.~D.~McLerran and B.~Svetitsky,
  Phys.\ Rev.\  D {\bf 24} (1981) 450.

\bibitem{Cel}
  T.~Celik, J.~Engels and H.~Satz,
  Phys.\ Lett.\  B {\bf 125} (1983) 411;
 R.~V.~Gavai,
  Nucl.\ Phys.\  B {\bf 633} (2002) 127
  [arXiv:hep-lat/0203015];
 S.~Datta and S.~Gupta,
  Phys.\ Rev.\  D {\bf 80} (2009) 114504
  [arXiv:0909.5591 [hep-lat]].


\bibitem{Gav}
  R.~V.~Gavai, F.~Karsch and H.~Satz,
  Nucl.\ Phys.\  B {\bf 220} (1983) 223.

\bibitem{DeT}
  C.~DeTar and U.~M.~Heller,
  Eur.\ Phys.\ J.\  A {\bf 41} (2009) 405
  [arXiv:0905.2949 [hep-lat]].

\bibitem{Pisa}
  R.~D.~Pisarski,
  Phys.\ Rev.\  D {\bf 74} (2006) 121703
  [arXiv:hep-ph/0608242].

\bibitem{Hel}
  U.~M.~Heller and F.~Karsch,
  Nucl.\ Phys.\  B {\bf 251} (1985) 254.


\bibitem{olaf1}
O.~Kaczmarek, F.~Karsch, P.~Petreczky and F.~Zantow,
  Phys.\ Lett.\  B {\bf 543} (2002) 41
  [arXiv:hep-lat/0207002].

\bibitem{Dumi}
A.~Dumitru, Y.~Hatta, J.~Lenaghan, K.~Orginos and R.~D.~Pisarski,
  Phys.\ Rev.\  D {\bf 70} (2004) 034511
  [arXiv:hep-th/0311223].


\bibitem{olaf2}
S.~Gupta, K.~Huebner and O.~Kaczmarek,
  Phys.\ Rev.\  D {\bf 77} (2008) 034503
  [arXiv:0711.2251 [hep-lat]].

\bibitem{whot}
T.~Umeda et al. , 
  Phys.\ Rev.\  D {\bf 79} (2009) 051501
  [arXiv:0809.2842 [hep-lat]].

\bibitem{whotlat08}
T.~Umeda et al. , 
  PoS\ LATTICE2008 (2008) 174
  [arXiv:0810.1570 [hep-lat]].

\bibitem{Eng}
  J.~Engels, J.~Fingberg and D.~E.~Miller,
  Nucl.\ Phys.\  B {\bf 387} (1992) 501.

\bibitem{Vel}
  A.~Velytsky,
  Int.\ J.\ Mod.\ Phys.\  C {\bf 19} (2008) 1079
  [arXiv:0711.0748 [hep-lat]].




\end{thebibliography}



\end{document}